\documentclass[aps,twocolumn,prl,superscriptaddress,preprintnumbers,showpacs,tightenlines]{revtex4}
\usepackage{amssymb}
\usepackage{epsfig,graphicx,times}

\begin{document}

\preprint{First Draft}
\title{Dressed Qubits in Nuclear Spin Baths}
\author{Lian-Ao Wu}
\affiliation{Department of Theoretical Physics and History of Science, The Basque Country
University (EHU/UPV), PO Box 644, 48080 Bilbao, Spain}
\affiliation{IKERBASQUE, Basque Foundation for Science, 48011 Bilbao, Spain}

\begin{abstract}
We present a method to encode a \textit{dressed} qubit into the product
state of an electron spin localized in quantum dot and its surrounding
nuclear spins via a dressing transformation. In this scheme, the hyperfine
coupling and a portion of nuclear dipole dipole interaction become logic
gates, while they are the sources of decoherence in electron spin qubit
proposals. We discuss errors and corrections for the dressed qubits.
Interestingly, the effective Hamiltonian of nuclear spins is equivalent to a
pairing Hamiltonian, which provides the microscopic mechanism to protect
dressed qubits against decoherence.
\end{abstract}

\maketitle

\textit{Introduction.---} The building blocks of quantum information
processors are controllable quantum bits. Electron spins in quantum dots are
promising candidates for these basic units \cite%
{Loss98,Burkard00,Kane98,Wu04}. The control of electron spins in quantum
dots has been investigated extensively in areas such as quantum information.
However, the decoherence, dominantly originating from the hyperfine coupling
between an electron spin and its surrounding nuclear spins in the host
material, may ruin the quantum process of the electron spin \cite%
{Khaetskii02,Merkulov02}. Distinct from random noise, the hyperfine coupling
causes \emph{inherent error} with non-Markovian feature \cite{Khaetskii02}
and can be manipulated to some extent \cite{Taylor03}. This nature has been
utilized to create long-lived quantum memory of electron spin qubits via the
surrounding nuclear spins and to implement optical pumping \ \cite%
{Taylor03,Imamoglu03,Taylor031,Kurucz09,Hu07,Cappellaro09}. Theoretical
motivations along this line have lead to interesting experimental results 
\cite{Churchill09,Jiang09,Lu09,Liu09}.

Alternatively, this inherent error may be corrected by the dressed qubit
method \cite{Wu03}. The essential ingredient to use this method is to find a
unitary dressing transformation between the basis of electron spin and the
product basis of electronic and nuclear spins, such that the matrix
representations of operators on the electron spin Hilbert space are the same
as those on the corresponding product space. This paper demonstrates the
feasibility of applying the dressed qubit method to the electronic-nuclear
spin system. Different from a bare electron spin interacting with nuclear
spins, the corresponding dressed qubit is only subject to leakage, which may
be suppressed by the Bang-Bang method in terms of a universal leakage
elimination operator \cite{Wu03}. Engineering of nuclear spin distribution
in the host material may also be an option in dealing with these leakages.
It is interesting to note that the effective Hamiltonian of nuclear spins is
equivalent to a pairing Hamiltonian, which helps the dressed qubit to
protect against decoherence.

\textit{Invariant subspaces spanned by electronic spin and nuclear spins.---}
Consider a single electron confined in quantum dot. The Hamiltonian for the
electron spin and its surrounding $K$ $(\approx 10^{5})$ nuclei with spin $I$
is 
\begin{equation}
H=H_{B}+H_{I}+H_{nuc},  \label{total}
\end{equation}%
where $H_{B}=g^{\ast }\mu _{B}BS_{z}+g_{n}\mu _{n}BI_{z}$ is the Zeeman
energy of the electron spin and nuclear spins in a magnetic field $B$ along
the \textit{z} axis. Here $S_{z}$ ( $I_{z}=\sum_{i}I_{z}^{i}$ ) is the 
\textit{z}-component of the electronic (total nuclear) spin operator. The 
\textit{z}-component of the total angular momentum, $J_{z}=S_{z}+$ $I_{z}$,
is conserved in the system. We can write the hyperfine coupling between
nuclear spins and the electron spin%
\begin{equation}
H_{I}=\mathcal{A}\sqrt{2I}(A_{z}S_{z}+V_{f}),  \label{coupling}
\end{equation}%
where $\mathcal{A}$ is an average hyperfine coupling constant. Operators $%
A_{\mu }=\sum_{i}\alpha _{i}I_{\mu }^{i}/\sqrt{2I}$ are expressed in terms
of the nuclear spin $I_{\mu }^{i}$ ( $\mu =z,+,-$ ), where the real numbers $%
\alpha _{i}$'s correspond to values of the electronic wave function at the
point $R_{i}$ and are normalized such that $\sum_{i=1}^{K}\alpha _{i}^{2}=1$
(sightly different from the normalization in Ref. \cite{Taylor03} ). The
dominant contribution of $A_{z}S_{z}$ is an effective magnetic field for the
electron spin, known as Overhauser shift \cite{Taylor03}. We will show later
that the effective magnetic field on the electron spin, including Overhauser
shift characterized by $\alpha _{i}$'s and $\mathcal{A}$, can be written%
\[
B_{eff}=B-\mathcal{A}\sum \alpha _{i}(I+\alpha _{i}^{2}/2)/g^{\ast }\mu _{B}.
\]%
The spin exchange $V_{f}=\frac{1}{2}A_{+}S_{-}+\frac{1}{2}A_{-}S_{+}$ plays
crucial roles in creating long-lived quantum memory \cite{Taylor03} and
implementing optical pumping \cite{Imamoglu03}. Significantly, this term
will also act as a logic gate in our scheme. The nuclear dipole dipole
interaction $H_{nuc}$ reads as%
\begin{equation}
H_{nuc}=%
\sum_{i=1;i<j}^{K}b_{ij}(I_{+}^{i}I_{-}^{j}+I_{-}^{i}I_{+}^{j}-4I_{z}^{i}I_{z}^{j}),
\label{dipole}
\end{equation}%
where $b_{ij}\propto (3\cos ^{2}\theta _{ij}-1)/r_{ij}^{3}$, $r_{ij}$ is the
distance between nuclei $i$ and $j$, $\theta _{ij}$ is the zenith angle of
the relative vector pointing from nucleus $i$ to $j$.

The nuclear spin operators may be represented in terms of fermionic pairs.
To each index $i$, we define a pair of \textquotedblleft imaginary state" $%
(i,\bar{\imath}),$ where $\bar{\imath}$ is the time reversal of the
imaginary state $i$. The nuclear spin operators $I_{-}^{i}$ and $I_{+}^{i}$
are then rewritten by fermionic pairs, 
\begin{equation}
I_{-}^{i}=\sum_{s=1}^{2I}c_{\bar{\imath}}^{s}c_{i}^{s}\text{, \ \ \ }%
I_{+}^{i}=\sum_{s=1}^{2I}c_{i}^{s\dagger }c_{\bar{\imath}}^{s\dagger },
\label{fermion}
\end{equation}%
which satisfy the restrictions $(I_{+}^{i})^{2I+1}=0.$ The commutator $%
[I_{-}^{i},I_{+}^{j}]=2\delta _{ij}(I-\hat{n}_{i})$ is represented by a
nuclear pair operator $\hat{n}_{i}=\sum_{\alpha =1}^{2I}(c_{i}^{s\dagger
}c_{i}^{s}+c_{\bar{\imath}}^{s\dagger }c_{\bar{\imath}}^{s})/2$. When $I=1/2$%
, the sums (\ref{fermion}) are simplified  $I_{-}^{i}=c_{\bar{\imath}}c_{i}$%
, \ \ \ $I_{+}^{i}=c_{i}^{\dagger }c_{\bar{\imath}}^{\dagger }$ and $\hat{n}%
_{i}=(c_{i}^{\dagger }c_{i}+c_{\bar{\imath}}^{\dagger }c_{\bar{\imath}})/2$.
A total nuclear pair operator can be defined as $\hat{n}=\sum_{i}^{K}\hat{n}%
_{i}$. Likewise, electron spin can be faithfully represented by pair
operators on a imaginary pair $(0,\bar{0})$ , $S_{-}=c_{\bar{0}}c_{0}$, $%
S_{+}=c_{0}^{\dagger }c_{\bar{0}}^{\dagger }$ and $S_{z}=\hat{n}_{0}-1/2$.
The total pair operator of the electron and nuclei is $\hat{N}=\hat{n}+\hat{n%
}_{0}$.

In recently proposed techniques of long-lived memory and optical pumping 
\cite{Taylor03,Imamoglu03}, it has been suggested that the dominant part of
the Hamiltonian ( \ref{total} ) is 
\begin{equation}
H_{D}=F(t)S_{z}+\mathcal{A}\sqrt{2I}V_{f},  \label{dominant}
\end{equation}%
where $F(t)=g^{\ast }\mu _{B}B_{eff}-g_{n}\mu _{n}B$ includes contributions
from electronic and nuclear spins as well as Overhauser shift. We have
neglected the constant $g_{n}\mu _{n}BJ_{z}$, where $J_{z}=\hat{N}-KI-1/2$ \
is conserved. There exist two-dimensional invariant subspaces of the
Hamiltonian $H_{D}$ for each given value of $N\in (0,2KI+1)$. In order to
show this explicitly, we consider a Hermitian operator $\hat{h}=A_{-}A_{+}$,
which commutes with the total nuclear pair operator $\hat{n}$. Let $%
\left\vert m\right\rangle $ be common eigenstates of the operators $\hat{h}$
and $\hat{n}$ such that $\hat{h}$ $\left\vert m\right\rangle
=h_{m}\left\vert m\right\rangle $. It is clear that the eigenvalues $%
h_{m}=\left\langle m\right\vert A_{-}A_{+}\left\vert m\right\rangle $ are
positive numbers. The two-dimensional subspaces, spanned by states 
\begin{equation}
\left\vert 0\right\rangle _{d}=\left\vert \uparrow \right\rangle
_{e}\left\vert m\right\rangle \text{, \ \ }\left\vert 1\right\rangle
_{d}=\left\vert \downarrow \right\rangle _{e}\left\vert \Phi
_{m+1}\right\rangle ,  \label{subspace}
\end{equation}%
are invariant under the Hamiltonian $H_{D}$. Here $\left\vert \uparrow
\right\rangle _{e}$ ( $\left\vert \downarrow \right\rangle _{e}$) is the
electron spin-up (down) state, and $\left\vert \Phi _{m+1}\right\rangle
=A_{+}\left\vert m\right\rangle /\sqrt{h_{m}}$ are nuclear spin states but
usually are not eigenstates of $\hat{h}$. $V_{f}$ exchanges the two states, 
\begin{eqnarray}
V_{f}\left\vert 0\right\rangle _{d} &=&\sqrt{h_{m}/4}\left\vert
1\right\rangle _{d}  \label{dress} \\
V_{f}\left\vert 1\right\rangle _{d} &=&\sqrt{h_{m}/4}\left\vert
0\right\rangle _{d}.  \nonumber
\end{eqnarray}%
Note that we have excluded two one-dimensional subspaces, where both
electronic and nuclear spins are completely polarized, with $N=0$ and $%
N=2KI+1$.

While there are many two-dimensional invariant subspaces characterized by
the total pair number $N$, we now concentrate on the $N=1$ invariant
subspace $\mathcal{H}_{2}$, which has been studied extensively. The
eigenstate $\left\vert m\right\rangle $ in this subspace is $\left\vert 
\mathbf{0}\right\rangle =\left\vert -I,-I,......,-I\right\rangle $ with
eigenvalue $h_{m}=1$, where nuclear spins are perfectly polarized. The state 
$\left\vert \Phi _{m+1}\right\rangle =A_{+}\left\vert \mathbf{0}%
\right\rangle $, denoted as $\left\vert \mathbf{1}\right\rangle $, is
orthogonal to the state $\left\vert \mathbf{0}\right\rangle $\textbf{\ }and
becomes an eigenstate of $\hat{h}$ in this particular case. In general,
given numbers $N$ and $I$, there are $\Omega (I,N)$ states in the \emph{%
combined} system of the electron spin and nuclear spins, for instance, when $%
I=1/2,$ $\Omega (1/2,N)=\frac{(K+1)!}{(K+1-N)!N!}$. The $N=1$ Hilbert space,
denoted by $\mathcal{H}_{K+1}$, is $K+1$- dimensional (i. e., $\Omega
(K,1)=K+1$). This means that there are additional $K-1$ states in the space,
which can be made orthogonal against the two states in eq. (\ref{subspace}).
The $K-1$ states are all in the electron spin-down state and can be written $%
\left\vert 1_{k}\right\rangle =\left\vert \downarrow \right\rangle
_{e}\left\vert \mathbf{1}_{k}\right\rangle $, where $\left\vert \mathbf{1}%
_{k}\right\rangle =A_{k+}\left\vert \mathbf{0}\right\rangle $ and $%
A_{k+}=\sum_{i}\alpha _{i}^{k}I_{+}^{i}/\sqrt{2I}.$ We identify the
"collective" mode $k=0$, i. e., $A_{+}=A_{0+}$ and $\alpha _{i}=\alpha
_{i}^{0}$. The set $\{\alpha _{i}^{k}\}$ corresponds to a $K\times K$ matrix 
$[\alpha ]$ and can, as usual, be made as a unitary matrix by using
Gram-Schmit orthogonalization such that $\left\langle \mathbf{1}_{k}|\mathbf{%
1}_{k^{\prime }}\right\rangle =\delta _{kk^{\prime }}$\cite{Kurucz09}. These
operators obey the commutation relations%
\begin{equation}
\lbrack A_{k-,}A_{k^{\prime }+}]=\delta _{kk^{\prime }}-\sum_{i}\alpha
_{i}^{k\ast }\alpha _{i}^{k^{\prime }}\hat{n}_{i}/I.  \label{commutor}
\end{equation}%
The Hilbert space $\mathcal{H}_{K+1}$ can be spanned by the orthogonal bases 
$\left\vert 0\right\rangle _{d}$, $\left\vert 1\right\rangle _{d}$ and $%
\left\vert 1_{k}\right\rangle $, where $k=1,...,K-1$. Note that with
equation ( \ref{commutor} ) we have $V_{f}\left\vert 1_{k}\right\rangle
_{l}=0,$ for all $k\neq 0.$

The bosonization of the nuclear spin operators has been used to discuss the
electron spin qubit protection against decoherence \cite{Kurucz09}. Consider
the bosonic form $V_{f}=\frac{1}{2}A^{\dagger }S_{-}+\frac{1}{2}AS_{+}$ of
the hyperfine coupling, where $A=\sum_{i}\alpha _{i}b_{i}$ corresponds to
the collective mode and $b_{i}$'s are bosons. The additional modes $%
A_{k}^{\dagger }=\sum_{i}\alpha _{i}^{k}b_{i}^{\dagger }$ are defined by
using the same matrix $[\alpha ]$ as the above. $A_{k}$ and $A_{k}^{\dagger }
$ obey the bosonic commutation relations%
\begin{equation}
\lbrack A_{k},A_{k^{\prime }}^{\dagger }]=\sum_{i}\alpha _{i}^{k\ast }\alpha
_{i}^{k^{\prime }}=\delta _{kk^{\prime }}.  \label{commutor1}
\end{equation}%
By comparison with Eq. (\ref{commutor}), it is clear that the nuclear spin
ensemble behaves like that of collective bosons when nuclear spins are in
well polarized states or $\sum_{i}\alpha _{i}^{2}n^{i}\ll I$.

\textit{Dressing transformation and single dressed qubit operations.---}
Here we introduce a \textit{dressing transformation} between the electron
spin space and the subspace $\mathcal{H}_{2}$, 
\[
W=\left\vert \uparrow \right\rangle _{e}\left\langle \uparrow \right\vert
\left\langle \mathbf{0}\right\vert +\left\vert \downarrow \right\rangle
_{e}\left\langle \downarrow \right\vert \left\langle \mathbf{1}\right\vert ,
\]%
which satisfies the unitary condition $WW^{\dagger }=W^{\dagger }W=1$ since $%
\left\vert 0\right\rangle _{d}\left\langle 0\right\vert +\left\vert
1\right\rangle _{d}\left\langle 1\right\vert =1$- the completeness in the
invariant subspace $\mathcal{H}_{2}$. Under this transformation $V_{f}$ acts
as $S_{x}:$ 
\begin{eqnarray*}
W^{\dagger }S_{x}W &=&\frac{1}{2}(\left\vert \uparrow \right\rangle
_{e}\left\langle \downarrow \right\vert \left\vert \mathbf{0}\right\rangle
\left\langle \mathbf{1}\right\vert +\left\vert \downarrow \right\rangle
_{e}\left\langle \uparrow \right\vert \left\vert \mathbf{1}\right\rangle
\left\langle \mathbf{0}\right\vert  \\
&=&\frac{1}{2}(\left\vert 0\right\rangle _{d}\left\langle 1\right\vert
+\left\vert 1\right\rangle _{d}\left\langle 0\right\vert )=[V_{f}].
\end{eqnarray*}%
In another word, the matrix representation of $V_{f}$ in $\mathcal{H}_{2}$
is the same as that of $S_{x}$ in the electronic spin space, i. e., 
\[
\lbrack V_{f}]=\frac{1}{2}\left( 
\begin{array}{cc}
0 & 1 \\ 
1 & 0%
\end{array}%
\right) =X_{d}/2,
\]%
denoted as $X_{d}/2$. Another operator $S_{z}$ is transformed as%
\[
W^{\dagger }S_{z}W=\frac{1}{2}(\left\vert 0\right\rangle _{d}\left\langle
0\right\vert -\left\vert 1\right\rangle _{d}\left\langle 1\right\vert
)=Z_{d}/2,
\]%
where $S_{z}$ in $\mathcal{H}_{2}$ plays the same role as that in the
electron spin space, denoted by $Z_{d}/2$. The Hamiltonian $H_{D}$ is
therefore rewritten 
\begin{equation}
H_{D}=F(t)Z_{d}/2+\mathcal{A}\sqrt{2I}X_{d}/2,  \label{universal}
\end{equation}%
and the dressed qubit is supported by the two states in (\ref{subspace}).
This form of the Hamiltonian, equivalent to that for the NMR quantum
computer, can generate a universal logic gate set for the dressed qubit,
even in cases when the hyperfine is not controllable. Single dressed qubit
gates can also be performed by using a sequence of square pulses, whose
evolution operators, in general, are%
\[
U(\phi ,\theta )=e^{-i\phi (\cos \theta Z_{d}+\sin \theta
X_{d})}=e^{-i\theta Y_{d}/2}e^{-i\phi Z_{d}}e^{i\theta Y_{d}/2}.
\]%
Here $\phi =t\sqrt{F^{2}+2I\mathcal{A}^{2}}>0$ and $\theta =\arctan (\sqrt{2I%
}\mathcal{A}/F).$ By controlling parameter $F$, we can manipulate the angles 
$\phi $ and $\theta .$ For instance, by setting $F=0$ we obtain $U(\phi ,\pi
/2)=e^{-i\phi X_{d}}$ and $U(\phi +\pi ,\pi /2)=e^{i\phi X_{d}}$. An
effective gate $Y_{d}=iZ_{d}X_{d}$ in $\mathcal{H}_{2}$ can be generated by
the circuit $U(\pi /2,\theta )X_{d}=ie^{-i(\theta +\pi /2)Y_{d}}$. We can
effectively generate any logic single-qubit operation in the invariant
subspace $\mathcal{H}_{2}$ with $e^{i\phi X_{d}}$ and $e^{-i\theta Y_{d}}$.

The same approach can be applied in the $N>1$ cases, except that another
nuclear spin eigenstate $\left\vert m\right\rangle $ of $\hat{h}$, other
than the perfectly polarized state, has to be initially prepared. Since the
perfectly polarized state usually is hard to be realized, it might be an
encouraging option to initially prepare another eigenstate instead, for
instance, a state with $I_{z}$ being zero.

\textit{Effective logic gates and leakage.---} Different from electron-spin
qubits, the present dressed qubits only suffers from leakage from the
dressed state $\left\vert 1\right\rangle _{d}$ into the rest of the Hilbert
space $\mathcal{H}_{K+1}$ spanned by $\left\vert 1_{k}\right\rangle $. The
leakage is caused by the residual effect of $A_{z}S_{z}$ ( or $A_{z}$, for $%
S_{z}\equiv -1/2$ in the leakage-related space) and the nuclear dipole
dipole interaction, which preserve the total nuclear pair number $n$.
However, it is interesting to note that, in the dressed qubit approach, the
major portion of the hyperfine term $A_{z}S_{z}$ and the dipole dipole
interaction only provide additional contributions to logic gates but do not
result in leakage.

As discussed previously, Overhauser shift is the major effect of $A_{z}S_{z}$%
. While $\left\vert 0\right\rangle _{d}$ is its eigenstate, i. e., $%
A_{z}\left\vert \mathbf{0}\right\rangle =c_{z}\left\vert \mathbf{0}%
\right\rangle $, the interaction $A_{z}S_{z}$ will ruin the state $%
\left\vert 1\right\rangle _{d}$ 
\[
A_{z}\left\vert \mathbf{1}\right\rangle =(c_{z}+\sum_{j}\alpha _{i}^{3}/%
\sqrt{2I})\left\vert \mathbf{1}\right\rangle +\left\vert O\right\rangle ,
\]%
where the constant $c_{z}=-\sqrt{I/2}\sum_{i=1}\alpha _{i}$. \ A part of the
second coefficient of the state $\left\vert \mathbf{1}\right\rangle $
contributes a constant in the subspace $\mathcal{H}_{2}$. The other part of
second coefficient and the first coefficient correspond to an additional
phase gate $-\mathcal{A}\sum \alpha _{i}(I+\alpha _{i}^{2}/2)Z_{d}/2$ -
Overhauser shift. The magnitude of the leaked state $\left\vert
O\right\rangle $ is much smaller than that of its orthogonal state $%
\left\vert \mathbf{1}\right\rangle $, with the relative ratio being
approximately $\frac{\alpha _{i}}{I\sum_{j}\alpha _{j}}\sim \frac{\mathcal{A}%
}{IK\mathcal{A}}\sim 10^{-5}$ . It is small but still in the order of the
fault tolerance threshold estimates of quantum error correction theory \cite%
{Steane02}.

Leakage also arises from the nuclear dipole dipole interaction. While $%
H_{nuc}\left\vert \mathbf{0}\right\rangle =c_{0}\left\vert \mathbf{0}%
\right\rangle $, the interaction acting on the other state yields $%
H_{nuc}\left\vert \mathbf{1}\right\rangle =(c_{0}+c_{1})\left\vert \mathbf{1}%
\right\rangle +\left\vert O^{\prime }\right\rangle $, where $\left\vert
O^{\prime }\right\rangle $ is a state orthogonal to $\left\vert \mathbf{1}%
\right\rangle $. The dominant contribution to $\mathcal{H}_{2}$ is $%
c_{0}=-16I^{2}\sum_{n<m}b_{nm}$, which is a constant in this subspace. The
second coefficient $c_{1}=4I\sum_{n\neq i}\alpha _{i}b_{ni}(8\alpha
_{i}+\alpha _{n})$ of the state $\left\vert \mathbf{1}\right\rangle $
indicates that the dipole dipole interaction also induces an additional
phase gate $c_{1}Z_{d}/2$ for the dressed qubit.

We now try to find a special form of the dipole dipole Hamiltonian that
preserves the subspace $\mathcal{H}_{2}$. Since the coupling constants $%
b_{ni}$ represent the classical dipole dipole interaction, the sum $%
\sum_{n}b_{ni}$ should stand for an average field acting on the $^{i}$th
nuclear spin due to all the others. We can assume that each spin is subject
to the same average field, i. e., $\sum_{n}b_{ni}=\bar{b}$ being a constant.
This assumption should be valid for homogeneous materials. We then consider
a family of $\left\{ b_{ni}\right\} $ satisfying the $K$ constraints $%
\sum_{i}b_{ni}\alpha _{i}=\tilde{b}\alpha _{n},$ where $\tilde{b}$ is a
constant. Based on the two assumptions, we can show $\bar{b}=\tilde{b}$ and
the dipole dipole interaction acts as%
\begin{equation}
(H_{nuc}-c_{0})\left\vert \mathbf{1}\right\rangle =36I\bar{b}\left\vert 
\mathbf{1}\right\rangle .  \label{cons}
\end{equation}%
This special form of $H_{nuc}$ does not cause leakage but provides
additional contribution to the phase gate. With this result, one may get rid
of leakage by adjusting the $K(K-1)/2$ coupling constants $b_{ij}$ towards
the $K$ constraints, as intimate as possible, through engineering the angles 
$\theta _{ij}$ and the distances $r_{ij}$.

The deviation from the special form causes leakage from the subspace $%
\mathcal{H}_{2}$ into the Hilbert space $\mathcal{H}_{K+1}$. We symbolize
the portion of Hamiltonian ( \ref{total} ) causing leakage as $H_{L}$, which
contains the leakage due to both $A_{z}S_{z}$ and $H_{nuc}$.

\bigskip \textit{Leakage elimination.--- }Leakage can be eliminated by
making use of fast \textquotedblleft bang-bang\textquotedblright\ pulses 
\cite{Wu02}. The key to this open-loop solution is to find a universal \emph{%
leakage-elimination operator} $R_{L}$ such that $R_{L}H_{L}R_{L}=-H_{L}.$
The leakage operator has the diagonal matrix representation in the space $%
\mathcal{H}_{K+1}$ 
\begin{equation}
\lbrack R_{L}]=\left( 
\begin{array}{cc}
-[I] & 0 \\ 
0 & [I^{\prime }]%
\end{array}%
\right) ,  \label{LEO}
\end{equation}%
where $-[I]$ is a $2\times 2$ unit matrix in dressed bases $\left\vert
0\right\rangle _{d}$ and $\left\vert 1\right\rangle _{d}$, and $[I^{\prime }]
$ is a $(K-1)\times (K-1)$ unit matrix in the rest of the space $\mathcal{H}%
_{K+1}$. \ It can be shown that the operator $R_{L}=\exp (-i\pi \lbrack
A_{+}S_{-}+A_{-}S_{+}])$ has the matrix representation ( \ref{LEO} ) and
thus is a leakage elimination operator. Leakage can be eliminated by the
standard bang-bang circuit $R_{L}\exp (-iH\tau /2)R_{L}\exp (-iH\tau /2)$ 
\cite{Wu02}, where time $\tau $ is made very short compared to the bath
correlation time. This circuit for the dressed qubit simplifies the error
control technique in electron spin qubits \cite{lidar09}.

\textit{Equivalent pairing Hamiltonian.--- }The hyperfine coupling induces
interaction among nuclear spins via the electron spin. An effective
correlation $V_{eff}=-\frac{\mathcal{A}^{2}I}{2F}A_{+}A_{-}$ can be
introduced by the well-known Fr\"{o}hlich transformation $e^{-S}V_{f}e^{S}$
with a generator $S=-\frac{\mathcal{A}}{F}\sqrt{I/2}(A_{-}S_{+}-A_{+}S_{-})$%
. The correlation is determined by $\mathcal{A}^{2}/F.$

By using Eq. (\ref{fermion}) and the induced nuclear interaction, we can
generically write the nuclear effective Hamiltonian (\ref{total}) as a
pairing Hamiltonian. To simplify, we consider the $I=1/2$ case, where the
nuclear effective Hamiltonian is%
\begin{equation}
H_{eff}=\sum_{i=1}^{K}\epsilon _{i}\hat{n}_{i}-2\sum_{i\neq j=1}^{K}b_{ij}%
\hat{n}_{i}\hat{n}_{j}-\sum_{i\neq j=1}^{K}g_{ij}c_{i}^{\dagger }c_{\bar{%
\imath}}^{\dagger }c_{\bar{j}}c_{j},  \label{spin1}
\end{equation}%
where $\epsilon _{i}=-\mathcal{A}\alpha _{i}/2-2\sum_{i\neq
j}(b_{ij}+b_{ji}) $ and $g_{ij}=\frac{\mathcal{A}^{2}}{4F}\alpha _{i}\alpha
_{j}+b_{ij}$. The first term corresponds to a signal particle energy of
imaginary states. The middle term stands for a on-site interaction and the
last is a standard pair correlation, where the dominant contribution stems
from the induced nuclear interaction $V_{eff}$. The ground state of the
effective Hamiltonian can be expressed approximately by the BCS wave
function, $\left\vert BCS\right\rangle \propto \exp (\sum_{k}\frac{%
v_{i}\alpha _{k}^{i\ast }}{u_{i}}A_{+k})\left\vert \mathbf{0}\right\rangle ,$
where $v_{i}$ and $u_{i}$ are obtained by solving the set of BCS\ equations
that can be found in textbooks (see, e. g., \cite{PRing}). The gap
parameters obey the self-consistent gap equations $\Delta _{i}=\frac{1}{2}%
\sum_{i}g_{ij}\Delta _{j}/\xi _{j}$, where $\xi _{j}=\sqrt{(\epsilon
_{j}-\lambda )^{2}+\Delta _{j}^{2}}$ and $\lambda $ is the chemical
potential determined by the nuclear pair number constraint $\left\langle
BCS\right\vert \hat{n}\left\vert BCS\right\rangle =n.$

Ref. \cite{Kurucz09} proposes a phenomenological scheme to protect the
nuclear spin memories by using the bosonization (\ref{commutor1}). The
scheme demonstrates that there is an energy gap between the collective
storage state, characterized by the collective boson $A$, and other states,
which plays the critical key to protect the quantum memory against local
spin-flip and spin-dephasing noise. Here the exact correspondence between
the nuclear spin Hamiltonian and the pairing Hamiltonian (\ref{spin1})
provides the microscopic mechanism of this energy gap and the scheme.

The set of BCS equations does not possess analytic solutions in the general
case. We can estimate the solution by setting $\alpha _{i}=1/\sqrt{K}$ and $%
b_{ij}=b.$ In this case, all $v_{i}$ are equal, $v_{i}=\sqrt{n/K},$\ \ \ $%
u_{i}=\sqrt{1-n/K}.$ The BCS wave function reads%
\[
\left\vert BCS\right\rangle \propto \exp (\sqrt{\frac{n}{K-n}}%
A_{+})\left\vert \mathbf{0}\right\rangle ,
\]%
where the nuclear dipole-dipole interaction does not contribute to the wave
function under this level of approximation. However, it appears in the gap
parameter%
\begin{equation}
\Delta =(\frac{\mathcal{A}^{2}}{4FK}+b)\sqrt{n(K-n)}.  \label{gap1}
\end{equation}%
The gap parameter keeps the BCS ground state away from other states. It also
indicates that when $n=K/2$ ( $I_{z}=0$ ) the gap reaches its maximum and
provides the most efficient protection for the BCS ground state against
decoherence. The result is only valid for fermionic pairs but not for
bosons.

\textit{Preparation, two-qubit gate and readout.--- }We now show that the
dressed qubits can be prepared and be read out. The preparation of the
polarized state $\left\vert \uparrow \right\rangle _{e}\left\vert
-I,-I,......,-I\right\rangle $ is the requirement for long-live quantum
memory \cite{Taylor03}. An optical technique has been proposed to achieve
the state \cite{Imamoglu03}. The idea is to utilize the hyperfine coupling
to induce the nuclear spin-flip process.

In their natural status, nuclear spins usually are in a mixed state with $%
N\approx K/2.$ It can be an option to distill the mixed state to initially
prepare another eigenstate of $\hat{h}$ , with $N$ being $(K\pm 1)/2$ or so,
providing that the polarization is too hard to be realized.

There are various versions of proposals for realization of controlled phase
gates between two spin qubits, for instance, by using Raman transitions
induced by classical laser fields \cite{Imamoglu99}. The two electron
correlation $S_{z}^{1}S_{z}^{2}$, generating the controlled phase gate for
spin qubits 1 and 2, can be translated directly into that of the dressed
qubits in the way that $Z_{d}^{1}Z_{d}^{2}/4=S_{z}^{1}S_{z}^{2}.$

Our dressed qubits can be read out directly through electron spins because
there is a one-to-one correspondence between dressed states and bare states
( \ref{subspace} ). The methods for spin-state measurements are available in
various proposals, e. g., ref. \cite{Loss98}.

\bigskip In conclusion, we have introduced a method to encode a \textit{%
dressed} qubit into an electron spin and nuclear spins. Unlike other
treatments against decoherence, the dressed qubit method does not require
extra overheads in gating, initialization and measurement. The hyperfine
coupling and a part of nuclear dipole dipole interaction now become logic
gates in this scheme, while they are sources of decoherence in electron spin
qubit proposals. The residual correlations from the hyperfine coupling $%
A_{z}S_{z}$ and dipole dipole interaction are categorized as leakages which
may be eliminated by the "Bang-Bang" method in a simple way. It is also
interesting to note a \emph{passive} strategy to reduce these leakages by
engineering the distribution of nuclear spins in the host material.

The author thanks Dr. W. Yao for helpful discussions. This work was
supported by the Ikerbasque Foundation.

\end{document}